\documentclass[conference]{IEEEtran}
\IEEEoverridecommandlockouts 

\usepackage{cite}

\ifCLASSINFOpdf
  \usepackage[pdftex]{graphicx}
\else
\fi
\usepackage{amsmath,amssymb,amsfonts}
\usepackage{algorithmic}
\usepackage{xcolor}

\newenvironment{ldescription}[1]
  {\begin{list}{}%
   {\renewcommand\makelabel[1]{##1\hfill}%
   \settowidth\labelwidth{\makelabel{#1}}%
   \setlength\leftmargin{\labelwidth}
   \addtolength\leftmargin{\labelsep}}}
  {\end{list}}
  
\usepackage{multirow}
\usepackage{balance}
\usepackage{comment}
\usepackage{eurosym}
\usepackage[mathscr]{euscript}

\begin{document}
%
\title{How Can Smart Buildings Be Price-Responsive?}

\author{\IEEEauthorblockN{Ricardo Fern\'andez-Blanco}
\IEEEauthorblockA{\textit{Group OASYS} \\
\textit{University of Malaga}\\
Malaga, Spain\\
Ricardo.FCarramolino@uma.es}
\and
\IEEEauthorblockN{Juan Miguel Morales}
\IEEEauthorblockA{\textit{Dept. Applied Mathematics} \\
\textit{University of Malaga}\\
Malaga, Spain \\
Juan.Morales@uma.es}
\and
\IEEEauthorblockN{Salvador Pineda}
\IEEEauthorblockA{\textit{Dept. Electrical Engineering} \\
\textit{University of Malaga}\\
Malaga, Spain \\
spinedamorente@gmail.com}

\thanks{This project has received funding from the European Research Council (ERC) under the European Union's Horizon 2020 research and innovation programme (grant agreement No 755705). This project has also been supported by Fundaci\'on Iberdrola Espa\~na 2018. The authors thankfully acknowledge the computer resources,
technical expertise and assistance provided by the SCBI
(Supercomputing and Bioinformatics) center of the University
of Malaga.} 
}

%


\maketitle

\begin{abstract}
The prospective participation of smart buildings in the electricity system is strongly related to the increasing active role of demand-side resources in the electrical grid. In addition, the growing penetration of smart meters and recent advances on home automation technologies will spur the development of new mathematical tools to help optimize the local resources of these buildings. Within this context, this paper first provides a comprehensive model to determine the electrical consumption of a single-zone household based on economic model predictive control. The goal of this problem is to minimize the electricity consumption cost while accounting for the heating dynamics of the building, smart home appliances, and comfort constraints. This paper then identifies and analyzes the key parameters responsible for the price-responsive behaviour of smart households.
\end{abstract}
\vspace{6pt}
\begin{IEEEkeywords}
Model predictive control, price-responsive loads, smart appliances, smart buildings.
\end{IEEEkeywords}

%
\IEEEpeerreviewmaketitle

\section*{Nomenclature}
A bar above ($\overline{\textcolor{white}{x}}$) or below ($\underline{\textcolor{white}{x}}$) any variable denotes its maximum and minimum bounds, respectively. 

\subsection{Sets and Indices}
\begin{ldescription}{$xxxx$}
\item [$\mathcal{C}_i$] Set of power phases of load $i$, indexed by $c$.
\item [$\mathscr{C}$] Set of user-defined comfort constraints.
\item [$\mathcal{I}$] Set of uninterruptible loads, indexed by $i$.
\item [$\mathcal{T}$] Set of time periods, indexed by $t$.
\item [$\mathcal{T}_i$] Set of desirable operation times of load $i$.
\item [$\mathcal{T}^{occ}$] Set of time periods during which the household is occupied, indexed by $t$.
\item [$\mathscr{T}$] Set of technical constraints pertaining to control actions.
\end{ldescription}\vspace{-2mm}
\subsection{Parameters}
\begin{ldescription}{$xxxxxx$}
\item [$CT_i$] Cycle duty of uninterruptible load $i$ [\# time periods].
\item [$N_T$] Number of time periods.
\item [$P^{ul}_{ci}$] Power of phase $c$ of uninterruptible load $i$ [kW].
\item [$P^{sb}_{t}$] Standby power consumption in period $t$ [kW].
\item [$TR_{it}$] Binary-valued parameter specifying the periods of operation of the uninterruptable load $i$.
\item [$UA^{r,a}$] Heat transfer coefficient between the room air and the ambient [W/$^\circ$C].
\item [$z^{amb}_{t}$] Outdoor ambient temperature in period $t$ [$^\circ$C].
\item [$z^{hwd}_{t}$] Hot water demand in period $t$ [l].
\item [$z^{occ}_{t}$] Occupancy schedule in period $t$.
\item [$\phi_t$] Luminance from natural light [lumen].
\item [$\eta^{le}$] Indoor luminous efficacy [lumen/kW].
\item [$\lambda_{t}$] Electricity price in period $t$ [\euro/kWh].
\item [$\rho^l$] Penalty term for the light level bounds [\euro/lumen].
\item [$\rho^{temp}$] Penalty term for temperature bounds [\euro/$^\circ$C].
\end{ldescription}\vspace{-2mm}

\subsection{Variables}
\begin{ldescription}{$xxxxxxx$}
\item [$l_{t}$] Light level in period $t$ [lumen]. 
\item [$u^b_{t}$] Power consumption of the building in period $t$ [kW].
\item [$u^{bl}_{t}$] Window blind position in period $t$ [p.u.].
\item [$u^{\imath}_{t}$] Power consumption in period $t$. Superscript $\imath$ = \{$heat$, $cool$, $hp$, $wh$, $rf$, $al$\} and they denote heating and cooling system, heat pump for the floor heating system, water heater, refrigerator, and artificial lighting, respectively [kW].
\item [$u^{ul}_{it}$] Power consumption of uninterruptible load $i$ in period $t$ [kW]. 
\item [$v^{l}_{t}$] Slack variable for the light level bounds in period $t$ [lumen].
\item [$v^{r/wh/rf}_{t}$] Slack variable for the temperature bounds of the room/water heater/refrigerator in period $t$ [$^\circ$C].
\item [$x^{\beta}_{t}$] Temperature in period $t$. Superscrit $\beta = \{ r, f, w, wh, rf \} $ and they denote temperature of room, floor, water in floor heating pipes, water heater, refrigeration chamber, respectively [$^\circ$C].
\item [$\delta^{a}_{cit}$] Auxiliary variable to relate power phase $c$ with the scheduling of uninterruptible load $i$ in period $t$.
\item [$\delta^{ul}_{it}$] Binary variable indicating whether uninterruptible load $i$ is scheduled on (1) or off (0) in period $t$.
\end{ldescription}\vspace{-2mm}

\subsection{Vectors}
\begin{ldescription}{$xxxxx$}
\item [$\boldsymbol{u}$] Vector of control variables. 
\item [$\boldsymbol{v}$] Vector of slack variables. 
\item [$\boldsymbol{x}$] Vector of state variables. 
\item [$\boldsymbol{y}$] Vector of measured signals. 
\item [$\boldsymbol{z}$] Vector of external disturbances. 
\end{ldescription}\vspace{-2mm}

\subsection{Matrices}
\begin{ldescription}{$xxxxx$}
\item [$\boldsymbol{A}$] Matrix of heat transfer coefficients (dynamics matrix). 
\item [$\boldsymbol{B}$] Matrix of coefficients relating the state and control variables (control matrix). 
\item [$\boldsymbol{C}$] Matrix of coefficients relating the state variables and the measured signals (sensor matrix). 
\item [$\boldsymbol{D}$] Matrix of coefficients relating the measured signals directly with the control variables (direct term).
\item [$\boldsymbol{E}$] Matrix of coefficients relating the state variables and external disturbances (disturbances matrix). 
\end{ldescription}

\section{Introduction}
Demand-side resources are expected to play an important role in future electricity systems \cite{Kirschen2003}. Their active participation in providing flexibility to the grid will also help integrate more and more renewable energy, thus paving the way towards the decarbonization of the electricity sector. Prospective demand-side resources are storage devices and smart buildings. Large-scale integration of the former is expected to increase provided capital cost reductions and increased charging/discharging efficiencies \cite{Dvorkin2017}, whereas the latter is gaining more attention due to recent advances on home automation and the growing penetration of smart meters in the distribution network \cite{COM2014}. This paper is focused on the smart buildings' ability to provide flexibility, which may be of utmost interest to incentivize the active participation of demand-side resources, thus leading to new ways of electricity trading \cite{Parag2016}.

There is a vast literature on building simulation by: (i) individually modelling thermal loads such as refrigerators \cite{Costanzo2013}, water heaters \cite{Sossan2013}, or smart solar tanks \cite{Halvgaard2012a}; or (ii) modelling the building heating dynamics \cite{Halvgaard2012}. Halvgaard \textit{et al.} \cite{Halvgaard2012} analyzed a single-zone building with a water-based floor heater (FH) via economic model predictive control (MPC) in order to shift power to periods with lower prices. Note that an economic MPC is a model that represents complex system dynamics while minimizing an economic objective function. 

The active participation of smart buildings in the electricity system has been widely investigated in the literature \cite{Li2016,Thomas2018,Saez-Gallego2018,Zhao2018,Contreras-Ocana2018,Anjos2018,Gonzalez2018}. Most of the works consider thermostatically-controlled loads (TCLs) only \cite{Thomas2018,Saez-Gallego2018,Zhao2018,Li2016}. Li \textit{et al.} \cite{Li2016} devised a market-based coordination of a pool of TCLs which were modelled by a second-order equivalent thermal parameter (ETP) model. Authors of \cite{Thomas2018} investigated the effects of dynamic-price retail contracts on integrated retail and power market operations wherein the households are represented by a simplified ETP model of their air conditioning systems. Authors of \cite{Saez-Gallego2018} simulated a pool of price-responsive households equipped with heating pumps to apply inverse optimization in order to forecast their consumption. Zhao and Zhang \cite{Zhao2018} developed a framework to aggregate price-responsive loads represented by a second-order ETP model. However, little attention has been paid to an integrated formulation of the smart building with smart appliances \cite{Anjos2018,Contreras-Ocana2018,Gonzalez2018}, which is crucial to accurately capture the potential flexibility of their buildings.

Anjos \textit{et al.} \cite{Anjos2018} proposed a Dantzig-Wolfe decomposition approach to deal with numerous and heterogeneous buildings managed by a central aggregator. Although, they model one of each load type that a building may have, the building heating dynamics were disregarded, thus ignoring the effect of the comfort constraints. Contreras-Oca\~na \textit{et al.} \cite{Contreras-Ocana2018} presented one of the most completed building models and analyzed possible interactions between commercial buildings and an aggregator of electric vehicles. Gonzalez \textit{et al.} \cite{Gonzalez2018} presented a residential load simulator with appliances. However, their focus was on providing load profiles rather than on how the household could provide flexibility. Recently, Junker \textit{et al.} \cite{junker2018} devised a flexibility metric to characterize generic buildings (or districts) that can measure their reactions to different penalty signals.

The major contributions of this paper are thus twofold: 

\begin{enumerate}
    \item We propose a compact formulation for modelling smart buildings with smart appliances, and a comprehensive formulation for a single-zone household. It includes a state-space model with five states in order to capture the building heat dynamics, comfort constraints with user-defined parameters, and technical constraints. As major novelties, the model includes the implementation of two different space heaters (water-based floor heater or HVAC systems) along with other thermal loads (e.g. refrigerator and/or water heater); and the uninterruptible loads are modelled with discrete variables to accurately account for their variable cycle power.
    \item We analyze the key drivers leading to price-responsive households based on their comfort settings and structural characteristics. 
\end{enumerate}


\section{Compact Problem Formulation}
\label{sec:compact}
Managing buildings with smart appliances call for an integrated operational model that takes into account the heating dynamics of the building. Economic MPC becomes thus a solid option to control the building and it has been applied to building dynamic simulation in the last decade (see \cite{Halvgaard2014} and references therein). This section provides a compact mathematical formulation for modelling a smart building by using economic MPC. The model can be mathematically expressed as the following optimization problem:
\begin{subequations}
\begin{align}
&\min_{\boldsymbol{u}, \boldsymbol{v}, \boldsymbol{x}, \boldsymbol{y}} \quad  \quad h(\boldsymbol{u}, \boldsymbol{v}) \label{general_fo}\\
&\text{subject to:} \quad \dot{\boldsymbol{x}} = \boldsymbol{A}\boldsymbol{x}+\boldsymbol{B}\boldsymbol{u}+\boldsymbol{E}\boldsymbol{z} \label{general_const1}\\
& \hspace{1.85cm} \boldsymbol{y} = \boldsymbol{C}\boldsymbol{x}+\boldsymbol{D}\boldsymbol{u} \label{general_const2}\\
& \hspace{1.85cm}\boldsymbol{u}, \boldsymbol{v}, \boldsymbol{y} \in \mathscr{C} \label{general_const3}\\
& \hspace{1.85cm} \boldsymbol{u} \in \mathscr{T}, \label{general_const4}
\end{align}
\end{subequations}
\noindent where $h(\cdot)$ comprises the electricity cost incurred by the smart building and the penalty cost due to discomfort, which is minimized in constraint \eqref{general_fo}. 

Constraints \eqref{general_const1}--\eqref{general_const2} define the state-space model of the building heat dynamics (including thermal loads), which are represented by a linear system of first-order differential equations. External disturbances can be accounted for in this system. Specifically, constraints \eqref{general_const1} model the heat balance in each building zone and thermal load (e.g. refrigerator or water heater). Constraints \eqref{general_const2} account for the relationship among the measured signals with the state and control variables. Equation \eqref{general_const3} models the set of user-defined comfort constraints, which include control over the indoor air temperature and lighting, among others. Finally, equation \eqref{general_const4} represents the set of technical constraints including the scheduling of uninterruptible loads such as washing machine, tumble dryer, dishwasher, smart oven, and so on.  
%

The system of differential equations can be discretized using Euler's method and thus the problem can be recast as:
\vspace{-0.2cm}
\begin{subequations}
\begin{align}
&\min_{\boldsymbol{u}, \boldsymbol{v}, \boldsymbol{x}, \boldsymbol{y}} \quad h(\boldsymbol{u}, \boldsymbol{v}) \label{general_fo_d}\\
&\text{subject to:}\notag\\
&\boldsymbol{x}_t = \boldsymbol{A}_d\boldsymbol{x}_{t-1}+\boldsymbol{B}_d\boldsymbol{u}_{t-1}+\boldsymbol{E}_d\boldsymbol{z}_{t-1}; \quad \forall t \in \mathcal{T} \label{general_const1_d}\\
&\boldsymbol{y}_t = \boldsymbol{C}_d\boldsymbol{x}_{t-1}+\boldsymbol{D}_d\boldsymbol{u}_{t-1}; \quad \forall t \in \mathcal{T}  \label{general_const2_d}\\
&\boldsymbol{u}, \boldsymbol{v}, \boldsymbol{y} \in \mathscr{C}  \label{general_const3_d}\\
&\boldsymbol{u} \in \mathscr{T}, \label{general_const4_d}
\end{align}
\end{subequations}
\noindent where expressions \eqref{general_fo_d}, \eqref{general_const3_d}--\eqref{general_const4_d} are identical to \eqref{general_fo}, \eqref{general_const3}--\eqref{general_const4}, in that order, but expressed in discrete time. Similarly, constraints \eqref{general_const1_d}--\eqref{general_const2_d} comprise the discrete state-space model described in \eqref{general_const1}--\eqref{general_const2}. Subscript $d$ denotes that the matrices are in discrete time.

\section{Price-Responsive Household Formulation}
\label{sec:household}
Given a single-zone household, we explore two options for space heating: 1) a water-based FH, and 2) an HVAC\footnote{FH stands for Floor Heater, while HVAC means Heating, Ventilation, and Air Conditioning system.} system. Apart from these thermal loads, we model a residential refrigerator and a water heater, as well as the indoor lighting. Uninterruptible loads are also represented in the model. Assuming that the state variables are directly observed (i.e., $\boldsymbol{C}_d$ and $\boldsymbol{D}_d$ are, respectively, equal to the identity and null matrices of appropriate dimensions), a price-responsive household can be formulated as:
\begin{align}
&\min_{\Xi}  \sum_{t \in \mathcal{T}} \lambda_{t} u^{b}_{t} + \sum_{t \in \mathcal{T}} \left[ \rho^{temp} \left( v^r_{t} + v^{wh}_{t} + v^{rf}_{t} \right) + \rho^l v^{l}_{t}  \right]
\label{obj}\\
&\text{subject to:}\notag\\
&\underbrace{\begin{pmatrix} x^r_{t} \\ x^f_{t} \\ x^w_{t} \\ x^{rf}_{t} \\ x^{wh}_{t} \end{pmatrix}}_{\boldsymbol{x}_t} = \boldsymbol{A}_d  \underbrace{\begin{pmatrix} x^r_{t-1} \\ x^f_{t-1} \\ x^w_{t-1} \\ x^{rf}_{t-1} \\ x^{wh}_{t-1} \end{pmatrix}}_{\boldsymbol{x}_{t-1}} +  \boldsymbol{B}_d \underbrace{\begin{pmatrix} u^{hp}_{t-1} \\ u^{heat}_{t-1} \\ u^{cool}_{t-1} \\ u^{al}_{t-1} \\ u^{bl}_{t-1} \\ u^{rf}_{t-1} \\ u^{wh}_{t-1} \end{pmatrix}}_{\boldsymbol{u}_{t-1}} + \boldsymbol{E}_d \underbrace{\begin{pmatrix} z^{amb}_{t-1} \\ z^{occ}_{t-1} \\ z^{hwd}_{t-1} \end{pmatrix}}_{\boldsymbol{z}_{t-1}}; \notag\\
& \hspace{7cm} \forall t \in \mathcal{T} \label{ss_model}\\
& \underline{x}^r_{t} - v_{t}^r \leq x^r_{t} \leq \overline{x}^r_{t} + v_{t}^r; \quad \forall t \in \mathcal{T} \label{comfort_room}\\
&\underline{x}^{wh}_{t} - v_{t}^{wh} \leq x^{wh}_{t} \leq \overline{x}^{wh}_{t} + v_{t}^{wh}; \quad \forall t \in \mathcal{T} \label{comfort_wh}\\
&\underline{x}^{rf}_{t} - v_{t}^{rf} \leq x^{rf}_{t} \leq \overline{x}^{rf}_{t} + v_{t}^{rf}; \quad \forall t \in \mathcal{T} \label{comfort_rf}\\
&\underline{l} - v^l_t \leq l_t \leq \overline{l} + v^l_t ; \quad \forall t \in \mathcal{T}^{occ}    \label{comfort_light}\\
&l_t = \phi_t u^{bl}_{t}  + \eta^{le} u^{al}_{t}; \quad \forall t \in \mathcal{T}^{occ}    \label{light_level}\\
&u^{bl}_{t} \geq \underline{u}^{bl}; \quad \forall t \in \mathcal{T}^{occ}    \label{comfort_blinds}\\
&u^{b}_{t} = \sum_{\imath} u^{\imath}_{t} + \sum_{i \in \mathcal{I}} u^{ul}_{it} + P^{sb}_{t}; \quad  \forall t \in \mathcal{T}    \label{total_power}
%
%
%
%
%
\end{align}

%
%
%
%
%
%

\begin{align}
%
%
%
%
%
&0 \leq u^{\imath}_{t} \leq \overline{u}^{\imath}; \forall \imath = \{heat, cool, hp, wh, rf, al\}, t \in \mathcal{T}     \label{lim_power} \\
& \delta^{ul}_{it} \leq TR_{it}; \quad \forall i \in \mathcal{I}, t \in \mathcal{T} \label{eq_dl0}\\
&\sum_{t \in \mathcal{T}}{\delta^{ul}_{it}} = CT_i; \quad \forall i \in \mathcal{I} \label{eq_dl1}\\
&\sum_{k = t}^{t + CT_i - 1}{\delta^{ul}_{ik}} \geq CT_i ( \delta^{ul}_{it} - \delta^{ul}_{i,t-1});  \forall t \leq N_{T} - CT_i + 1 \label{eq_dl2}\\
&\delta^{a}_{1it} \geq \delta^{ul}_{it} - \delta^{ul}_{i,t-1}; \quad \forall i \in \mathcal{I}, t \in \mathcal{T} \label{eq_dl4}\\
&\delta^{a}_{cit} \geq \delta^{a}_{c-1,i,t-1}; \quad \forall c > 1,  i \in \mathcal{I}, t \in \mathcal{T} \label{eq_dl5}\\
&u^{ul}_{it} = \sum_{c \in \mathcal{C}}{\delta^{a}_{cit} P^{ul}_{ci}}; \quad \forall i \in \mathcal{I}, t \in \mathcal{T}  \label{eq_dl6}\\
& \delta^{a}_{cit} \in [0, 1]; \quad \forall c \in \mathcal{C},  i \in \mathcal{I}, t \in \mathcal{T} \label{eq_bin_delta_act}\\
& \delta^{ul}_{it} \in \{0, 1\}; \quad \forall i \in \mathcal{I}, t \in \mathcal{T} \label{eq_bin_delta_dl}\\
& u^{bl}_{t} \in [0, 1]; \quad \forall t \in \mathcal{T} \label{eq_bin_delta_bl}\\
& \underbrace{\begin{pmatrix} v^r_{t}, v^{rf}_{t},  v^{wh}_{t},  v^l_t \end{pmatrix} ^T}_{\boldsymbol{v}_t} \geq \boldsymbol{0}; \quad \forall t \in \mathcal{T}, \label{eq_nonnegative_v}
\end{align}

\noindent where $\Xi = \left(l_t, \boldsymbol{x}_t, \boldsymbol{u}_t, u_t^b, \boldsymbol{v}_t, \delta^a_{cit}, \delta^{ul}_{it}\right)$  is the set of variables.

The goal of this optimization problem, given in \eqref{obj}, is the minimization of the electricity costs and penalty costs on violations of user-defined comfort constraints. We assume that a set of prices is known \textit{a priori} by the household's energy management system. The state-space model of the household heat dynamics is given by constraints \eqref{ss_model}, whose matrices are described in \cite{Fernandez-Blanco2018}. This model includes a three-state model for the water-based FH, which is represented by the inside air temperature, the floor temperature, and the water temperature in the floor heating pipes, as similarly done in \cite{Halvgaard2012}. In addition, the state-space model \eqref{ss_model} takes into account two additional states, i.e., the thermal dynamics of a residential refrigerator and a water heater located outside the household. The interested reader is referred to \cite{Costanzo2013, Sossan2013} for complex models on these two appliances. Moreover, the HVAC system is also represented in the matrix equations given in \eqref{ss_model} by a single-state model (only the indoor air temperature). The external disturbances are the ambient temperature, room occupancy, and hot water demand.

Comfort constraints are set in \eqref{comfort_room}--\eqref{comfort_blinds}, expressions \eqref{total_power}--\eqref{eq_dl6} represent technical constraints, whereas \eqref{eq_bin_delta_act}--\eqref{eq_nonnegative_v} declare the character of variables $\delta^{a}_{cit}$, $\delta^{ul}_{it}$, $u^{bl}_{t}$, and the slack variables $v^r_{t}$, $v^{rf}_{t}$, $v^{wh}_{t}$, $v^{l}_{t}$.

Regarding the comfort constraints, \eqref{comfort_room}--\eqref{comfort_rf} set the user-defined minimum and maximum bounds on the indoor air temperature, water temperature in the water heater, and air temperature in the refrigerator chamber, in that order, for each time period $t$. The comfort bounds for the indoor air temperature can be chosen as: $\overline{x}^r_t = x^{r,set}_t + \alpha w_t$ and $\underline{x}^r_t = x^{r,set}_t - \alpha w_t$, where $x^{r,set}_t$ is the set-point indoor air temperature, $\alpha$ is the maximum temperature difference with respect to the set-point that the user is able to withstand, and $w_t$ is a vector of continuous parameters varying between 0 and 1. Thus, comfort bounds were set up in two ways:  

\begin{itemize}
    \item \emph{Price-independent comfort bounds} (PI-CB): $w_t$ is a vector of ones.
    \item \emph{Price-dependent comfort bounds} (PD-CB): $w_t$ is a vector of normalized prices over a given day.
\end{itemize}

Constraints \eqref{comfort_light}--\eqref{comfort_blinds} model the household light levels when occupants are in the household, as described in \cite{Contreras-Ocana2018}. Constraints \eqref{comfort_light} enforce recommended minimum and maximum light levels. The household light level given in \eqref{light_level} can be computed as the natural light coming through the windows, which depends on the blind positions (i.e., $\phi_t u^{bl}_{t}$), and the indoor artificial lighting (i.e., $\eta^{le} u^{al}_{t}$). Constraints \eqref{comfort_blinds} may enforce a lower bound for the position of the blinds. 

Regarding the technical constraints, expression \eqref{total_power} sets the building electricity consumption equal to the contributions from the heat pump of the FH, HVAC system, indoor artificial lighting, water heater, refrigerator, uninterruptible loads, and stand-by power. Constraints \eqref{lim_power} impose the bounds on the power for the heating pump of the FH, HVAC system, artificial lighting, water heater, and refrigerator. The set of constraints \eqref{eq_dl0}--\eqref{eq_dl6} defines the operation of the uninterruptible loads. We assume that each uninterruptible load is characterized by its cycle time $CT_i$ and the cycle power at each phase $c$, i.e., $P^{ul}_{ci}$. In addition, the user may pre-define the time interval $\mathcal{T}_i$ at which the load could be scheduled on so that the binary-valued parameter $TR_{it} = 1$ if the uninterruptible load $i$ could be on in period $t$, and $TR_{it} = 0$ otherwise. Expressions \eqref{eq_dl0} enforce the uninterruptible loads to be scheduled off outside the pre-defined time interval, whereas constraints \eqref{eq_dl1} enforce binary variables $\delta_{it}^{ul}$ to be $1$ during the cycle time within the pre-defined time interval. Constraints \eqref{eq_dl2} impose that the scheduling during the cycle time must be consecutive. Constraints \eqref{eq_dl4}--\eqref{eq_dl5} set the relationship between the scheduling variable $\delta_{it}^{ul}$ and the activation variable $\delta_{cit}^{a}$. Finally, expressions \eqref{eq_dl6} define the power of the uninterruptible load $i$ in period $t$ as the power in the corresponding cycle phase.


\section{Case Study}
\label{sec:case}
The single-zone household presented in \cite{Halvgaard2012} is used to analyze what comfort settings and structural attributes will help make it price-responsive. The floor and window areas are respectively 30 and 1 m$^2$. For the water-based FH, the data are based on \cite{Halvgaard2012}. The mass of water in the FH is 400 kg. The compressor has a coefficient of performance (COP) of 3 and a nominal power of 1 kW. When the household is equipped with an HVAC system, the COP for the heating and cooling system is respectively 1.67 and 3.67 and its electrical capacity is 1 kW. The household is equipped with a 30-l water heater located outside and a residential refrigerator. Data can be found in Table \ref{tab1:data_rf_wh}. Inlet water temperature and the hot water daily consumption profile is given in \cite{Fernandez-Blanco2018}. 

\begin{table}[tbp]
\caption{Data for the Refrigerator \cite{Costanzo2013, Gonzalez2018} and Water Heater \cite{Sossan2013}}
\vspace{-0.3cm}
\begin{center}
\begin{tabular}{|l|c|c|}
\cline{2-3}
\multicolumn{1}{c|}{ } & \textbf{Refrigerator} & \textbf{Water heater}\\
\hline
Nominal power capacity [kW]& 0.35 & 1.26 \\
COP & 0.76 & 0.92\\
Thermal capacity [Wh/$^\circ$C] & 6.65& 34.85\\
Heat transfer coefficient [W/$^\circ$C] & 0.678 & 0.5 \\
\hline
\end{tabular}
\label{tab1:data_rf_wh}
\end{center}
\end{table}

\begin{table}[tbp]
\caption{Data for the Comfort Constraints}
\vspace{-0.3cm}
\begin{center}
\begin{tabular}{|l|l|c|c|c|}
\cline{3-5}
\multicolumn{1}{c}{ } & \multicolumn{1}{c|}{ } & \textbf{\emph{noflex}} & \textbf{\emph{flex}}& \textbf{\emph{extraflex}}\\
\hline
\multicolumn{1}{|c}{ }&$x^{r,set}_t$ [$^\circ$C] & 20 & 20 & 20 \\
\multicolumn{1}{|c}{ }&$\alpha$ [$^\circ$C]& 0 & 2 & 5\\
\hline
\multicolumn{1}{|c}{ }&\{$\underline{x}^{wh}$, $\overline{x}^{wh}$\} [$^\circ$C] & \{54, 56\} & \{50, 60\} & \{45, 65\} \\
\hline
\multicolumn{1}{|c}{ }&\{$\underline{x}^{rf}$, $\overline{x}^{rf}$\} [$^\circ$C]& \{4.9, 5.1\} & \{4, 5\} & \{3, 6\}\\
\hline
\multirow{5}{*}{$\mathcal{T}_i$} & Washing machine & \multicolumn{3}{c|}{06:00-14:00} \\
& Dishwasher (first) & \multicolumn{3}{c|}{06:00-14:00}\\
& Dishwasher (second) & \multicolumn{3}{c|}{16:00-00:00}\\
& Tumble dryer & \multicolumn{3}{c|}{15:00-00:00}\\
& Oven & \multicolumn{3}{c|}{10:00-15:00}\\
\hline
\end{tabular}
\label{tab2:data_comfort}
\end{center}
\end{table}

\begin{table*}[htbp]
\caption{Results for a Single-Zone Household}
\vspace{-0.3cm}
\begin{center}
\begin{tabular}{|c|c|c|c|c|c|c|c|c|c|c|}
\cline{1-11}
\multirow{2}{*}{\textbf{Type}} & \textbf{Comfort} & \multirow{2}{*}{\textbf{Case}} &  \textbf{Annual} & \textbf{Violations} &  \textbf{Freq. at} & \textbf{Freq. [18,20)--} & \textbf{Freq. [15,18)--} &  \textbf{Building} &  $u^{b}$ & $u^{hp/heat}$ \\
 & \textbf{bounds} & {} &  \textbf{cost} [\euro] & [$^\circ$C $\cdot$ h]&  \textbf{20$^\circ$C} [\%]$^{\mathrm{a}}$ & \textbf{(20,22] $^\circ$C} [\%]$^{\mathrm{a}}$ &  \textbf{(22,25]$^\circ$C} [\%]$^{\mathrm{a}}$ &   \textbf{cons.} [kWh] &  [\%]$^{\mathrm{b}}$ &  [\%]$^{\mathrm{b}}$ \\
\hline
\multirow{6}{*}{FH} & \multirow{3}{*}{PI-CB} & \emph{noflex}    &             103.5 &           862.5 &             37.5 &                          62.5 &                           0.0 &             1944.0 &            54.5 &                  51.3\\
&  & \emph{flex}      &             78.9 &              0.0 &              1.5 &                          98.5 &                           0.0 &             1569.1 &            62.7 &                  86.6 \\
&  & \emph{extraflex}      &             68.4 &              0.0 &              0.0 &                           0.3 &                          99.7 &             1372.7 &            63.5 &                  85.8\\
\cline{2-11}
& \multirow{3}{*}{PD-CB} & \emph{noflex}    &             103.5 &           862.5 &             37.5 &                          62.5 &                           0.0 &             1944.0 &            54.5 &                  51.3 \\
&  & \emph{flex}    &             93.2 &            55.9 &             22.2 &                          77.8 &                           0.0 &             1843.9 &            61.9 &                  70.0 \\
&  & \emph{extraflex}    &             87.0 &            35.2 &             21.7 &                          78.3 &                           0.0 &             1751.3 &            64.1 &                  73.1 \\
\cline{1-11}
\multirow{6}{*}{HVAC} & \multirow{3}{*}{PI-CB} & \emph{noflex}    &            107.0 &              0.0 &            100.0 &                           0.0 &                           0.0 &             2044.1 &            57.2 &                  60.2 \\
&  & \emph{flex}    &              89.9 &              0.0 &              1.1 &                          98.9 &                           0.0 &             1793.0 &            63.5 &                  77.2 \\
&  & \emph{extraflex}    &            76.1 &              0.0 &              0.5 &                          19.6 &                          79.9 &             1538.8 &            64.7 &                  80.1 \\
\cline{2-11}
 & \multirow{3}{*}{PD-CB} & \emph{noflex}    &            107.0 &              0.0 &            100.0 &                           0.0 &                           0.0 &             2044.1 &            57.2 &                  60.2 \\
&  & \emph{flex}    &             94.5 &              0.0 &             11.1 &                          88.9 &                           0.0 &             1867.7 &            62.3 &                  70.6  \\
&  & \emph{extraflex}    &             84.2 &              0.0 &              8.0 &                          61.6 &                          30.4 &             1691.2 &            64.4 &                  73.9 \\
\hline
\multicolumn{10}{l}{$^{\mathrm{a}}$Columns 6--8 represent the percentage of 15-minute time intervals lying within the given temperature intervals throughout the year.} \\
\multicolumn{10}{l}{$^{\mathrm{b}}$Columns 10--11 represent the share of power lying within low-price periods.}
\end{tabular}
\label{tab1:results_base_case}
\end{center}
\end{table*}

Artificial lighting capacity is 60 W with an indoor luminous efficacy equal to 90 lumen/W. We assume that the luminous efficacy of daylight is 105 lumen/W in order to compute the outdoor illuminance in lux and the comfort light levels are set to 100 and 10000 lux. Note that the outdoor illuminance and the comfort light levels are respectively multiplied by the window and floor area to convert them to lumen.  

We consider 4 uninterruptible loads: oven, washing machine, tumble dryer, and dishwasher. Their schedules are described in \cite{Fernandez-Blanco2018}, their desirable operating hours are given in Table \ref{tab2:data_comfort}, and their power on each cycle phase can be found in \cite{Stamminger2008} for year 2020. 

The discretization step $\Delta t$ is assumed to be 15 min to properly capture the building dynamics and the penalty terms $\rho^{temp}$ and $\rho^l$ are set to 1000. We run daily simulations with 15-min time steps for one year. For each simulation, we use a look-ahead window of one day to account for the future impacts of thermal dynamics when control actions are taken in a given period of time. Ambient temperature, solar radiation, and electricity prices are given in \cite{Fernandez-Blanco2018} for the sake of reproducibility. This reference also includes the occupancy schedules. Note that the standby power is neglected.

Three different cases regarding the degree of household flexibility are analyzed: \emph{noflex}, \emph{flex}, and \emph{extraflex} cases. Table \ref{tab2:data_comfort} also provides the data on setting comfort bounds for the indoor air, refrigerator, and water heater temperature.

The simulations have been performed on a Linux-based server with one CPU clocking at 2.6 GHz and 2 GB of RAM using CPLEX 12.6.3 \cite{cplex} under Pyomo 5.2 \cite{pyomo}. Optimality gap is set to 0\%.

\subsection{Effect of Comfort Settings}
\label{sub:effect_comfort}
Table \ref{tab1:results_base_case} provides results for a single-zone household under two types of space heating (FH and HVAC), three cases for the degree of flexibility (\emph{noflex}, \emph{flex}, and \emph{extraflex}), and the strategy followed to set the comfort bounds (PI-CB and PD-CB). This table shows the annual electricity cost, the total violations (i.e., number of Celsius degrees out of the corresponding comfort bounds over the 35040 time periods of the year), the percentage of 15-minute time intervals in which the indoor air temperature lies within three different ranges, the building energy consumption, and the share of building and heating power lying in low-price periods (i.e., those periods where the annual normalized prices are lower than 0.5). Note that we show the frequency of time intervals in which the indoor air temperature lies within: (i) 20.0 $^\circ$C (column 6), (ii) [18, 20)--(20, 22] $^\circ$C (column 7), and (iii) [15, 18)--(22, 25] $^\circ$C (column 8). For all cases, computing times for running annual simulations were in the range of 960--3600 s. 

The HVAC system leads to more expensive solutions than the FH, but provides in general less discomfort assuming that the reference set-point of 20$^\circ$C is the temperature at which the occupants experience the highest comfort. Note that for the \emph{noflex} case, the HVAC is able to keep the air temperature within comfort bounds (i.e., at the reference set-point), unlike a household with FH (this is also due to the absence of air conditioning when considering FH). We can notice that the annual cost decreases when increasing the degree of flexibility at the expense of a higher degree of discomfort regardless of the comfort settings. Also, the cost reduction is higher under the PI-CB strategy when increasing the degree of flexibility  (33.9\% vs. 15.9\% for the \emph{extraflex} case with respect to the \emph{noflex} case with the FH). Similar conclusions can be drawn with the HVAC system. However the PD-CB strategy leads to higher costs than under the PI-CB when the occupants are flexible regardless of the type of space heater.

The PD-CB strategy results in time-varying comfort bounds throughout the day, which is the reason why it leads to higher costs than the PI-CB. However, we can clearly observe that the PD-CB leads to less discomfort (assuming that the reference set-point of 20$^\circ$C is the temperature at which the occupants experience the highest comfort). Note that the percentage of 15-minute time intervals lying within [15, 18)--(22, 25] $^\circ$C is significantly reduced under the PD-CB compared to that percentage under the PI-CB.

Columns 10--11 in Table \ref{tab1:results_base_case} provide the share of power (total in the building and that of the space heating system) lying in low-price periods. This would be an indicator whether a household is more or less price-responsive. Percentage of total power (column 10) in low-price periods increases as the occupants are more flexible. This trend becomes even more noticeable when it comes to the heating power since space heaters have a higher degree of control than other loads such as refrigerators, dishwashers, etc. Both space heating systems allow for shifting energy to low-price periods when increasing the flexibility (around 7.2--7.5\% of difference between the share of building power for the \emph{extraflex} and \emph{noflex} cases with HVAC versus 9.0--9.6\% with FH). Although there are some differences, both systems can be price-responsive when using a look-ahead window.


\begin{table}[tbp]
\caption{Effect of $UA^{r,a}$ on Annual Cost and Discomfort Metrics}
\vspace{-0.4cm}
\begin{center}
\begin{tabular}{|c|c|c|c|c|c|}
 \hline
 \multirow{2}{*}{\textbf{Type}} &    \multirow{2}{*}{\textbf{Cost and metrics}}      &        \multicolumn{4}{c|}{\textbf{Factor}}\\
 \cline{3-6}
 &        &        \textbf{\textit{0.5}} &         \textbf{\textit{1}} &         \textbf{\textit{2}} &        \textbf{\textit{4}}\\
\hline
\multirow{5}{*}{FH}&Cost [\euro]              &       76.6 &        86 &     105.8 &     149.8 \\
&Violations [$^\circ$C $\cdot$ h]              &      0 &         0 &       1.1 &     93.4 \\
&Freq. at 20$^\circ$C [\%]   &        0 &         0 &       2.9 &       4.2 \\
 &{Freq. [18,20)--(20,22]}$^\circ$C [\%] &       100 &       100 &        97 &      95.4 \\
 &{Freq. [15,18)--(22,25]}$^\circ$C [\%] &        0 &         0 &       0.1 &       0.2 \\
\hline
\multirow{5}{*}{HVAC}&Cost [\euro]              &       87 &     106.5 &       146 &       225\\
&Violations [$^\circ$C $\cdot$ h]              &      0 &         0 &         0 &         0  \\
&Freq. at 20$^\circ$C [\%]   &        0.9 &       0.6 &       0.3 &       0.2  \\
& {Freq. [18,20)--(20,22]}$^\circ$C [\%] &       99.1 &      99.4 &      99.7 &      99.8 \\
& {Freq. [15,18)--(22,25]}$^\circ$C [\%] &        0 &         0 &         0 &         0 \\
 \hline
\end{tabular}
\label{tab3:results_UA1}
\end{center}
\end{table}

\subsection{Effect of Structural Parameters}
\label{sub:effect_structural}
Structural parameters such as the heat transfer coefficients\footnote{The heat transfer coefficient $UA$ is the product of the heat conductivity $U$ and the surface area $A$ where the heat transfer takes place.} could be also responsible for the household price-responsiveness. Table \ref{tab3:results_UA1} shows the effect of the heat transfer coefficient between the room air and the ambient, $UA^{r,a}$, for the \emph{flex} case on the annual cost, total number of violations of comfort bounds, and the percentage of 15-minute time intervals in which the room temperature lies within three different ranges, whereas Table \ref{tab3:results_UA2} shows the share of building power\footnote{Building power refers to the total consumption of the smart household.} lying within low-price periods for different flexibility cases. In order to isolate the effect of the heat transfer coefficient, we assume that the household does not have any window and that the lighting constraints are ignored. The value of $UA^{r,a}$ has been multiplied by factors of 0.5, 1, 2, and 4, respectively. The greater the value of $UA^{r,a}$, the less insulated the household is. 

It can be observed than when the household is less insulated, the annual costs increase because there are more thermal exchanges through the walls at the expense of slightly increasing the occupants' discomfort (see Table \ref{tab3:results_UA1}). Since this table shows the results from the \emph{flex} case, most of the time periods are above 18$^\circ$C and below 22$^\circ$C.

\begin{table}[tbp]
\caption{Effect of $UA^{r,a}$ on the Share of Building Power Consumption Lying within Low-price Periods [\%]}
\vspace{-0.4cm}
\begin{center}
\begin{tabular}{|c|c|c|c|c|c|c|}
\cline{2-7}
\multicolumn{1}{c|}{} &  \multicolumn{3}{c|}{\textbf{FH}} &  \multicolumn{3}{c|}{\textbf{HVAC}} \\
\hline
\textbf{Factor} &  \textbf{\emph{noflex}} &  \textbf{\emph{flex}} &  \textbf{\emph{extraflex}} &  \textbf{\emph{noflex}} & \textbf{\emph{flex}} &  \textbf{\emph{extraflex}} \\
\hline
0.5 &   54.9 &  64.2 &      65.5 &   56.1 &  63.0 &      65.2 \\
1.0 &   54.0 &  67.2 &      68.1 &   55.9 &  62.3 &      64.6 \\
2.0 &   53.1 &  71.0 &      71.7 &   55.7 &  60.2 &      62.2 \\
4.0 &   51.7 &  72.5 &      73.4 &   55.6 &  57.8 &      58.6 \\
\hline
\end{tabular}
\label{tab3:results_UA2}
\end{center}
\end{table}

The household with an FH is more price-responsive when increasing the value of $UA^{r,a}$ (see Table \ref{tab3:results_UA2}). We can observe that the differences between the \emph{extraflex} and \emph{noflex} cases vary between 10.6--21.7\% for the factors 0.5--4, respectively. Conversely, the household with an HVAC system is more price-responsive when decreasing the value of $UA^{r,a}$. This is due to the fact that the HVAC system is an appliance with fast dynamics whereas the water-based FH is characterized by slow dynamics.

\section{Conclusions}
\label{sec:conclusion}
We propose a compact and integrated formulation for a smart building with smart appliances via economic model predictive control. 
We model five states to capture thermal dynamics of space heaters, refrigerator and water heater; and we accurately model variable power cycles for uninterruptible loads.

Substantial cost savings can be achieved when increasing the comfort bounds under HVAC systems (21--29\%) and water-based floor heating systems (16--34\%). Price-responsiveness of smart households is quite similar between space heaters of slow (water-based FH) or fast impact (HVAC system) if a look-ahead window is used. The amount of building consumption lying within low-price periods increases from 54.5--57.2\% to 64.1--64.7\% when using wider comfort bounds for the smart appliances. In general, the more price-responsive the household is, the higher discomfort the occupants experience. However, price-dependent comfort bounds could help reduce the occupants' discomfort, thus reducing the number of time intervals at temperatures far from the reference set-point. Insulated households with HVAC systems are more price-responsive and they may lead to cost savings up to 50\% approximately, whereas price-responsiveness in households with water-based FHs is higher for less insulated ones at the expense of increasing the annual costs.  


Further research will be devoted to exploring the use of battery energy storage systems and local renewable generation to increase the flexibility degree of smart buildings. 

\balance

\bibliographystyle{ieeetr}
\bibliography{powertech2019}

\end{document}